\documentclass[runningheads]{llncs}
\usepackage[T1]{fontenc}
\usepackage{graphicx}
\usepackage{hyperref}
\begin{document}
\title {Let a million entrepreneurs grow!}
%
%

\author{Mrityunjay Kumar\inst{}\orcidID{0000-0003-2819-759X}}
\authorrunning{Kumar}
\institute{Independent Consultant at Palash \\
\email{mkumar@palash.com}
}



%
\maketitle              
\begin{abstract}
India produces about nine hundred thousand (900K) engineers annually, and many seek computer science and related technology jobs. Given that the IT workforce in India is still young, new graduates get jobs only when the industry grows. A liberal estimate based on the data from MeitY (Ministry of Electronics and Information
Technology) and NASSCOM puts the annual job growth to three hundred thousand (300K), less than one-third of the graduation rate. In other words, about half a million graduates don't get a job every year (even when we consider that some students don't opt for jobs or go for higher studies). 

This position paper demonstrates that given the current growth rate of the Indian economy, such a significant shortfall will continue to exist. It then proposes a way to address this shortfall. 

The paper proposes to develop micro-entrepreneurs at scale, enabling many graduates to start micro-enterprises focused on AI, Software, and Technology (MAST). These MAST enterprises offer technology products and services to meet the hyperlocal needs of the businesses and individuals in the local community (a retailer in the neighborhood, a high net-worth person, or a factory).

Such an endeavor will require curricular, policy, and societal interventions. The paper presents an approach to enable MAST education across campuses, outlining the key curricular changes required and important policies that must be created and implemented. 

This supply-demand gap is an existential problem for engineering education in India, and this position paper aims to trigger debates and collaborations to devise solutions that will work at India scale.


\end{abstract}

%
%
%

\section{Introduction}
\label{sec:introduction}
India graduates over 900K engineers in a year \cite{aishe2022}. Most of these engineers look for IT and allied services jobs, given the industry's past performance in generating jobs and the traditional lack of core engineering jobs. 

The IT sector includes four entities: a) Indian IT services companies (TCS, Infosys, Wipro, etc.), b) India centers of global services companies (Accenture, IBM, Capgemini, etc.), c) Global Capability Centers, or GCCs (captive technology centers of large multinational companies that serve the parent company), and d) Tech startups and products companies. Tech startups and products are the smallest categories, though growing quite rapidly \cite{indiasoftwareproductecosystem2022,indiantechecosystem2024}, and are likely to drive technology-driven growth. 

The IT sector employed 5.43 million (direct employment) in 2023-24 \footnote{\href{https://www.meity.gov.in/content/employment}{www.meity.gov.in/content/employment}}, but hired only 60K in 2023-24 (a sharp decline from the years before). There is a stark difference between the number of students graduating every year (900K) and the number employed last year (60K). Even the peak hiring of COVID times (an anomaly) barely employed 50\%  of the new graduates. 

The graduating rate (supply) far exceeding the headcount creation rate (demand) will likely stay this way for some time (see Figure \ref{fig:itsector-projectedheadcount}). The magnitude of this problem (large-scale unemployment of engineers) is so high that it is no longer just a higher education problem but a social one. As a society, we must find other ways to help engineering graduates gain employment, given that the IT sector is unlikely to do so. 

We believe \textbf{micro-entrepreneurship} is how to address this at scale. A micro-entrepreneur is an entrepreneur on a small scale - minimal capital investment, hyperlocal customers, a very small setup, and a low-risk environment. In India, micro-enterprises comprise more than 97\% of the MSME (Micro, Small, and Medium Enterprises). They are usually discussed as a poverty-alleviation scheme in developing countries \cite{hussain2014entrepreneurship}. However, we believe they effectively address the supply-demand gap of engineers and propose adapting this to suit the needs of engineering graduates who want to earn after graduating. 

Entrepreneurship has been proposed as a way to restructure engineering education \cite{venkateswarlu2017establishing}. Models of entrepreneurship education in engineering have also been studied and proposed \cite{da2015entrepreneurship}. The Entrepreneurial Intention of students in higher education in India has been studied \cite{pandit2018examining} as well, and the researchers found that explicitly teaching entrepreneurship does increase the intent. They also found STEM students possess skills that are already suitable for entrepreneurship. Therefore, leveraging entrepreneurship to address the supply-demand gap at scale is not completely novel and can yield results if used suitably. We intend to show the suitability in this position paper. 

The paper makes a case for enabling an ecosystem where we equip the students to be micro-entrepreneurs so that they can run micro-enterprises after graduation and not rely on finding a job to be gainfully employed. We outline why this is necessary and then suggest concrete ideas and approaches to achieve this at scale. 

Section \ref{sec:demand} reviews the data available on the demand side (number of jobs being created annually) and shows the projections based on the data. We rely primarily on MeitY (Ministry of Electronics and Information Technology) and NASSCOM data. Section \ref{sec:supply} reviews the data available on the supply side (engineering students graduating annually). Section \ref{sec:supply-demand-gap} highlights the gap and proposes micro-entrepreneurship as a solution. Section \ref{sec:micro-entrepreneurs-scale} describes how micro-entrepreneurs can be developed at scale on the campuses through curricular, policy, and societal interventions and introduces MAST as the specialized form of micro-enterprises we envisage. Section \ref{sec:mast-enterprises-ecosystem} presents ideas on how we can enable and support the MAST enterprises when they run their enterprises. Section \ref{sec:challenges-risks-mitigations} details the practicalities and other concerns that might be raised against the proposal. Section \ref{sec:conclusion-futurework} concludes by reviewing how the solution proposed in this paper can produce an impact, outlines a few challenges to address, and reiterates the need to solve this massive problem we face.

\section{Demand side}
\label{sec:demand}
\begin{figure}
    \centering
    \includegraphics[width=0.8\linewidth]{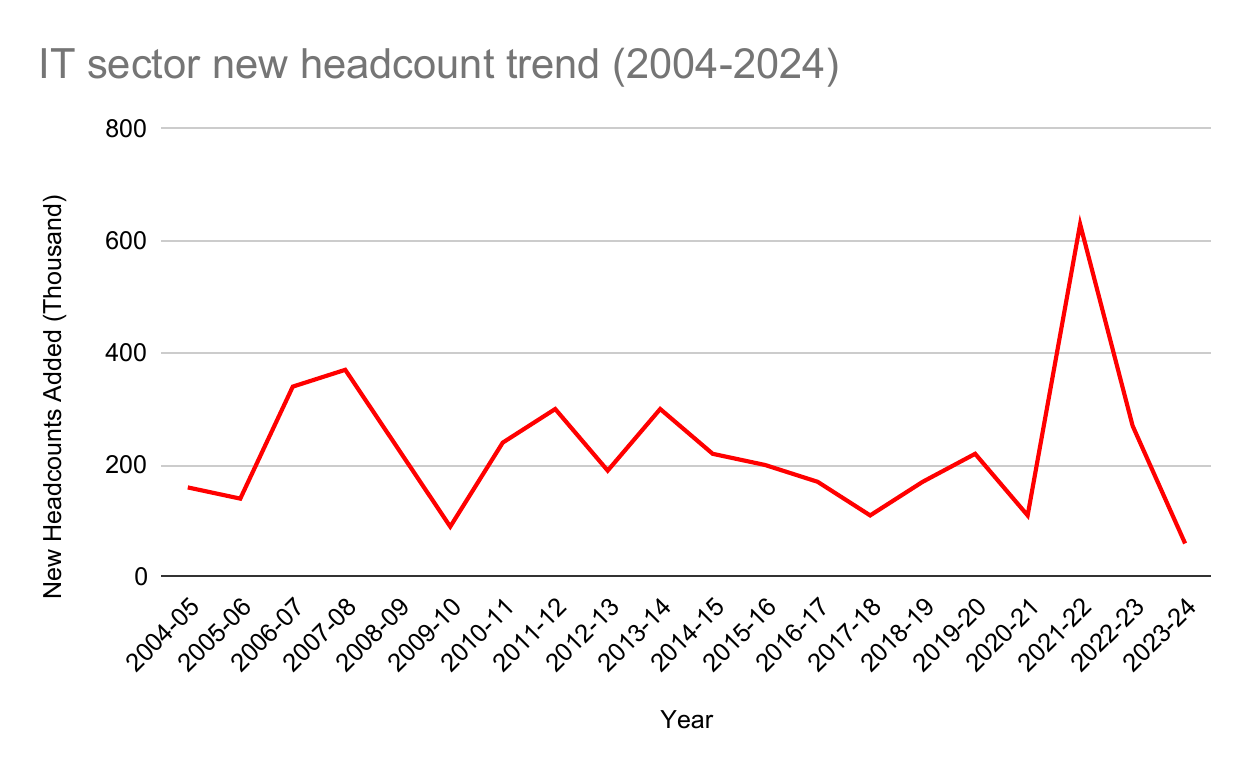}
    \caption{New headcounts added yearly over the last twenty years}
    \label{fig:new-headcounts}
\end{figure}

According to the data from MeitY (Ministry of Electronics and Information Technology) shown in Figure \ref{fig:new-headcounts}, new headcounts created year on year have been fluctuating, peaking during the COVID period (445K) and sliding to an all-time low (60K) in 2023-24. 

\begin{figure}
    \centering
    \includegraphics[width=0.9\linewidth]{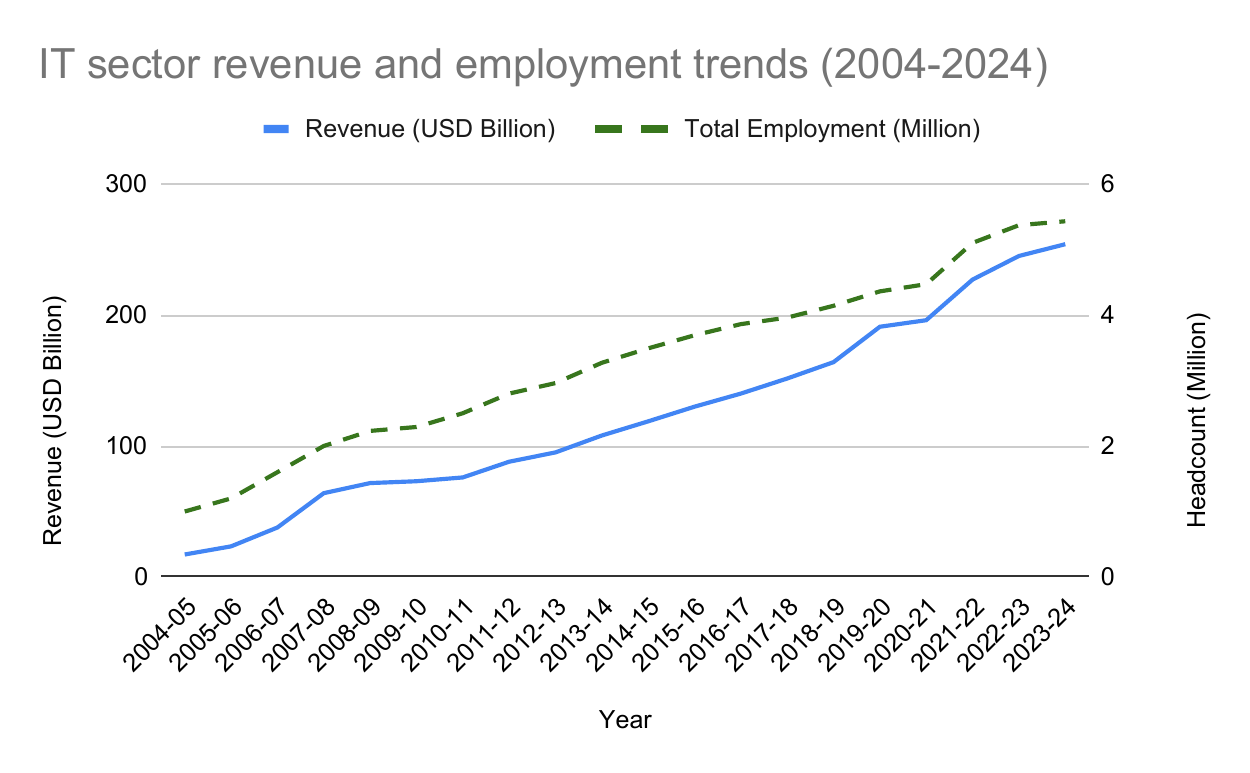}
    \caption{IT Sector revenue and headcount trends over the last twenty years}
    \label{fig:itsector-revenue-headcount}
\end{figure}

The new headcount is dependent on how the revenue grows. The revenue growth over the last five years (Figure \ref{fig:itsector-revenue-headcount}) of the sector had a CAGR (Compound Annual Growth Rate) of 5.87\%. The headcount growth has been much more erratic (Figure \ref{fig:new-headcounts}) with a similar rate. 

\begin{figure}
    \centering
    \includegraphics[width=0.9\linewidth]{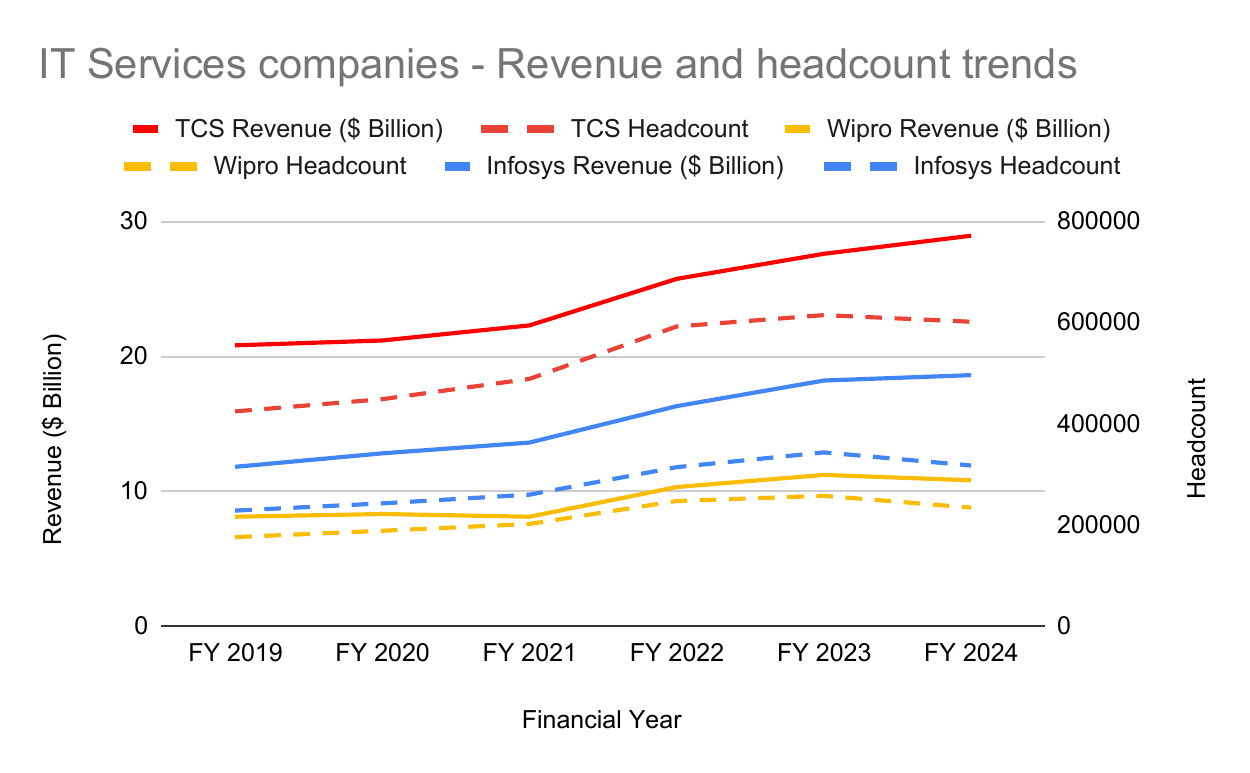}
    \caption{IT Services companies - Revenue and headcount trends}
    \label{fig:itservices-revenue-headcount}
\end{figure}

Figure \ref{fig:itsector-revenue-headcount} shows the sector revenue and headcount over the last five years\footnote{\href{https://www.meity.gov.in/content/software-and-services-sector}{www.meity.gov.in/content/software-and-services-sector}} and Figure \ref{fig:itservices-revenue-headcount} shows the same for three large IT services companies\footnote{Company annual reports}. 

It can be seen that the revenue and the headcount track each other closely and have an almost linear relationship. IT companies track this 'revenue per employee' metric (R/E metric) yearly and try to improve it.  

\subsection{Projected growth rate of headcount 2025-2030}
We want to use existing data and estimates to project the new headcounts created over the next five years to see how the demand will change. To do so, we will assume a revenue growth rate and apply a factor using the revenue per employee metric to derive the headcount needs for each year. Based on the 2022-23 values, the R/E metric is 45K USD. For our calculation, we interpret it to mean that for every 45K increase in revenue, a new headcount is created in the sector. 

We take three growth rate scenarios for projection: Current Rate (6\%), GDP Rate (7\%), and Optimistic Rate (8\%).

\begin{figure}
    \centering
    \includegraphics[width=0.9\linewidth]{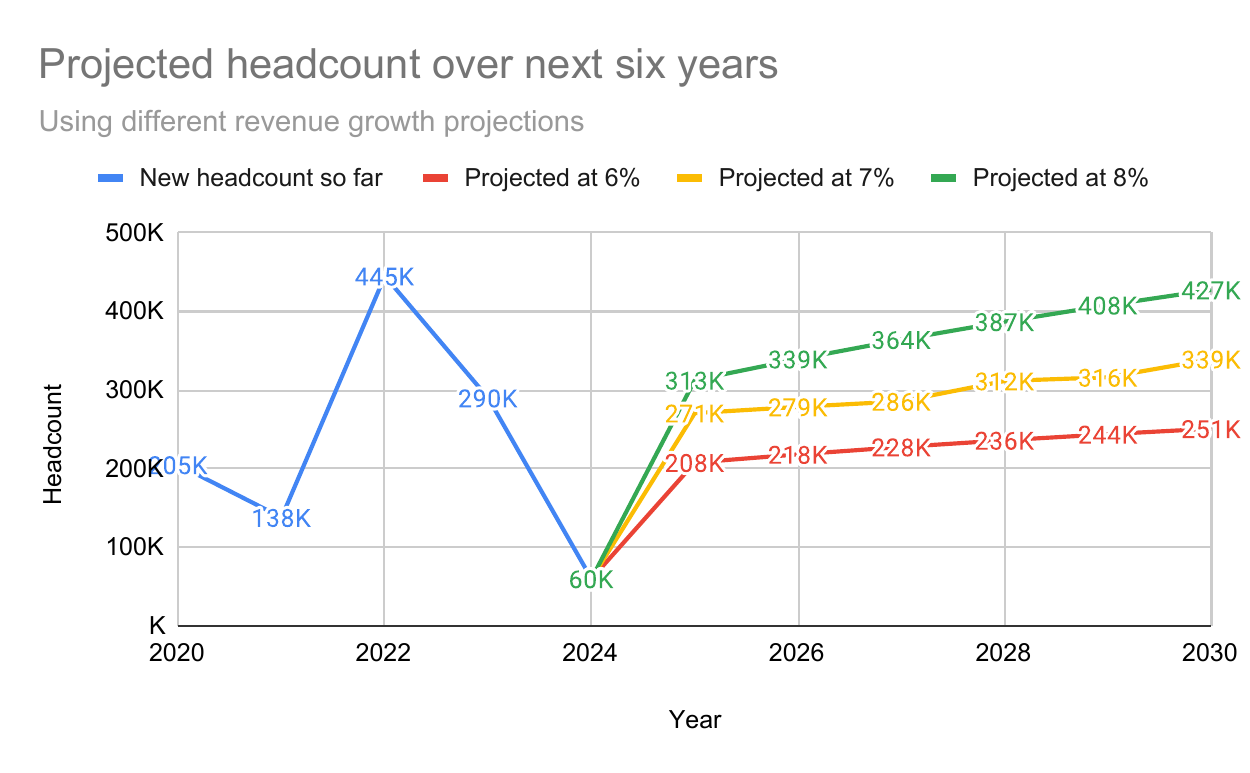}
    \caption{Projected headcount over next six years, with three different projected revenue growth rates: current (6\%), GDP rate (7\%), an Optimistic rate (8\%) }
    \label{fig:itsector-projectedheadcount}
\end{figure}

There is a strong push by IT companies to leverage AI and automation and improve employee productivity to generate more revenue through fewer engineers. This is a strong headwind that will reduce the headcount creation going forward. To partially account for this, the numbers in Figure \ref{fig:itsector-projectedheadcount} assume an increase in R/E by 2\% every year (current value is 1.5\%). This may be an underestimation.  

Figure \ref{fig:itsector-projectedheadcount} shows the new headcounts that will be added over the next six years under these three growth rate scenarios. Even at the optimistic rate projection, we will have less than 50\% of the required headcounts (900K) by 2030. This is a dire state of affairs.

\section{The supply side}
\label{sec:supply}
India graduates more than 900K engineers in a year, about a third of which are in computer science (CS) and related disciplines (300K) \cite{aishe2022}.

\begin{figure}
    \centering
    \includegraphics[width=0.9\linewidth]{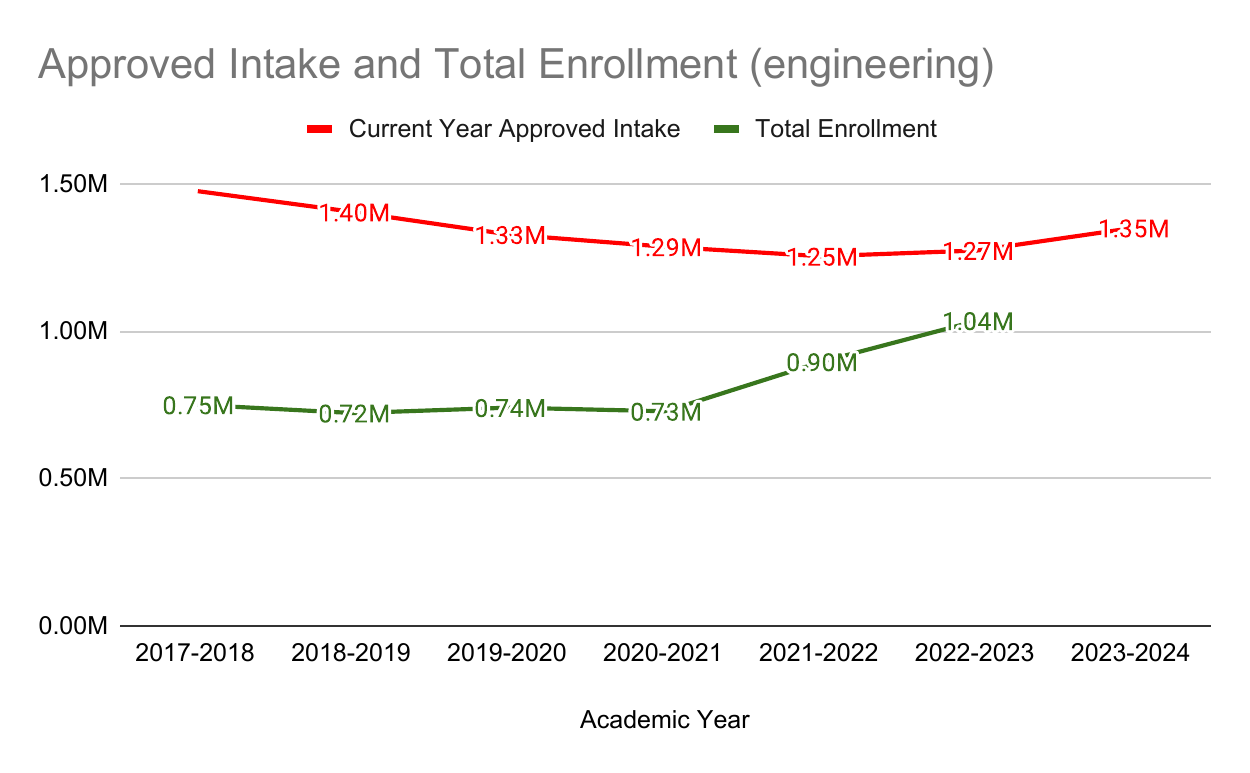}
    \caption{Intake approved from AICTE vs. Total enrollment in engineering UG over the years}
    \label{fig:intake-enrollment-UG}
\end{figure}

Figure \ref{fig:intake-enrollment-UG} demonstrates the cyclic nature of approved intake and enrollment numbers, which causes the bullwhip or Forrester effect \cite{forrester1997industrial}. This is caused by feedback loops (with delays) between approved intake and enrollment and between enrollment and placement records. A big positive change in any quantity (for example, extremely high hiring in COVID times) causes a ripple through the system that will last several years and push the supply up, which has been the case with IT over these two decades. 

\textbf{Why do we have so many students enrolling for an engineering course?}
One reason for this rush to engineering courses is that IT jobs are considered lucrative compared to other jobs. AICTE publishes data for other professional courses it regulates \cite{aicte2024}.
The enrollments for non-engineering degrees have stayed constant over the last 7-8 years. Interestingly, diploma and UG enrollments have shrunk while PG enrollments have increased. 
This may also suggest an inflation of degree in non-engineering programs - you need to be a post-graduate to do most of the jobs (and hence spend 5-6 years before getting employed). This can also drive people to engineering courses, which can provide employment after four years of education.  

The severity of the supply-demand mismatch mentioned above is not apparent to the student's parents and well-wishers, who continue believing in an IT course's lucrativeness based on a small number of success stories. The top 15 and bottom 15 engineering colleges (excluding IITs and NITs) of the top 100 NIRF-ranked colleges (2023 ranking) have average placement rates of 84\% and 69\% \cite{NIRF2023}, respectively. This falls precipitously for lower-ranked colleges, given the overall sectoral lack of jobs. These placement records cause the entire sector to be considered lucrative, and enrollments remain robust. This is further exacerbated by the coaching industry (with more than 70K crore business in India \cite{IndiaToday2020}), which has a significant stake in engineering entrance test preparation. When students invest so much to prepare for top engineering colleges, they enroll in mediocre colleges when they can't get into their desired college (sunk cost fallacy\footnote{\href{https://en.wikipedia.org/wiki/Sunk\_cost}{https://en.wikipedia.org/wiki/Sunk\_cost}}). 

\textbf{What about the quality of these engineering graduates? }
The quality of the education these graduates receive is a cause for concern. Some data suggest that the graduates do not possess enough skills to be employed even in entry-level roles of an engineer. Some estimates \cite{graduateskillindex2023} put this number at 47\%. In other words, out of the 900K engineering graduates, less than 450K are employable. However, even if we assume conservatively that only 400K are employable, the 2023-24 hiring (60K) was way less than this number, so even the employable engineers did not get jobs. 

While one reason for the lack of employable skills in 500K (unemployable) graduates is teaching of insufficient quality, we believe another reason is the diversity of students who enroll. Since every science student tries to get an engineering degree, colleges get many students who do not have the aptitude or interest in core engineering skills and end up graduating with poor engineering skills. Given the poor diversity in the jobs available to these graduates (since everyone wants to get a typical IT job), many otherwise suitable graduates are deemed unemployable.

\section{Addressing supply-demand gap}
\label{sec:supply-demand-gap}
As shown in the previous section, the gap in the number of engineering graduates and jobs will be more than 500K \textit{every year} and will continue to increase. This paints a grim picture for the foreseeable future. 

One way to address this is to drive up the demand; another way is to drive down the supply. While we need to do both, these require complex solutions that will take a long time to take effect.  

This paper proposes a solution that addresses this gap: Entrepreneurship. We propose to train the students to set up and operate small businesses focused on local needs. This serves two purposes: it reduces the supply of engineers who seek jobs right after graduation, and it positively impacts the local economy (which has the potential to increase jobs and prosperity at the local level).  

This is technology entrepreneurship but with a difference. 

Technology entrepreneurship is typically associated with high-risk and high-reward endeavors, where the goal is to produce a large impact at a global level. We call it Model A. Campuses worldwide try to create an environment that fosters entrepreneurial culture and develops student entrepreneurs using Model A \cite{marchand2014student}, and Venture Capitalists (VCs) fund these endeavors with the hope of generating outsized returns on their investments. These attributes make typical technology entrepreneurship focus on global impact and high-risk behavior (chasing growth rather than profitability, for example). Model A is a high-risk, high-reward, global-impact entrepreneurship. 

We propose producing entrepreneurs who operate in a low-risk environment, produce stable returns, and work at the local level. We call it Model B. Model B requires very low capital and instead relies on the labor and intellectual capital provided by the entrepreneurs; the focus is always on self-sufficiency and revenue generation. 

Model A behavior is not sustainable when we want to produce hundreds of thousands of entrepreneurs. We propose producing micro-entrepreneurs who can set up and operate Model B entrepreneurs and micro-enterprises.  

MSMEs (Micro, Small and Medium enterprises) provide 62\% of employment in India \cite{mgimsme2024}, and more than 95\% of MSMEs are micro-enterprises (most of which are in the manufacturing and services industries). Using size to distinguish between entrepreneurial organizations \cite{audretsch2012entrepreneurship}, our Model B enterprises fall in the category of micro-enterprises, albeit in a technology domain. We expect these engineering students to become micro-entrepreneurs and set up and operate micro-enterprises.

To distinguish current micro-enterprises from the new technology-driven micro-enterprises we envisage, we will call the latter MAST (\textbf{M}icro-enterprises focused on \textbf{A}I, \textbf{S}oftware and \textbf{T}echnology) enterprises.  

The MAST entrepreneurs invest time and effort, and the need for capital will be very small. They work with local businesses and individuals to offer services at a zero or negligible marketing cost. One of the key differentiators these micro-entrepreneurs can offer, compared to similar offerings by a national or international provider, is exemplary customer service because they are just a call away. The other differentiator is their ability to understand the hyperlocal context and better serve their customers. For example, while a local business can use \href{https://www.wix.com/}{Wix}\footnote{https://www.wix.com/} to create their website, a MAST enterprise can help the business by understanding their needs better and being available to help them build and use the website effectively. MAST enterprises can cater to needs like digitization, business process improvement, automation, and innovation in business verticals like real estate, agriculture, govt service deliveries, etc. 
 
MAST enterprises will be technology enterprises and not just an IT services setup; IT services and product offerings could provide a way to start their business.  When they are successful, MAST enterprises can generate demand for engineers (not only computer engineers) and help improve the demand side of the supply-demand gap as well. When these entrepreneurs sustain their enterprise for a few years, they gain valuable skills that can fetch them jobs should they need them. Hundreds of thousands of MAST enterprises foster an entrepreneurial culture and a Make in India drive, increasing the probability of creating blockbuster technology startups from India. 

We envisage producing 500K MAST entrepreneurs yearly across all the engineering campuses in India. The next sections provide details on how such a scale can be achieved.

\section{Producing MAST entrepreneurs at scale}
\label{sec:micro-entrepreneurs-scale}
We need to enable engineering students to develop into MAST entrepreneurs. This is proposed to be achieved through curriculum, policy, and societal interventions. 

\subsection{Curricular interventions}

A MAST enterprise needs the founders to possess diverse skills to handle all aspects of the business - business development, product and service creation, customer support and success, and overall business management and growth. An individual micro-entrepreneur is unlikely to have all the skills, but a group of entrepreneurs can possess all the required skills together. 

Therefore, all teaching and learning will happen in cohorts created based on interest and prior skills. We call this a MAST cohort, which consists of five members with complementary skills and operates together as a high-performance team \cite{katzenbach2015wisdom}. A new cohort is formed at the beginning of each year and stays together for the academic year. 

A MAST cohort gets hands-on experience running an enterprise through the creative use of course projects and assignments. These experiences are designed in a way that students get to deal with real-world problems through case studies \cite{raju1999teaching} and other means, get immediate feedback on their performance, get to repeat their experiences multiple times, and perform deliberate practice \cite{ericsson2015acquisition}. These are designed to ensure expertise can be developed. The students engage in apprenticeship with active MAST enterprises (maybe alumni of the same college) to practice their skills in the real world. 

\subsection{Policy interventions}
Curriculum changes require policy interventions. MAST education (skills and experiences required to acquire MAST skills) must be part of the engineering curriculum, and appropriate credit rationalization must be done. We can set up MAST as a specialization or minor in engineering disciplines, and students can opt for MAST credits starting in their first year. 

Internships and major projects should be allowed to be a part of MAST credits. Financial support during internships and major projects for these students (technology building, travel and stay, access to labs, etc.) should be made available. 

We also need policy changes to encourage and incentivize colleges to support MAST education and enable students to become MAST entrepreneurs. Some existing policies relevant to skill development and entrepreneurship \cite{economicsurvey2024} should be aligned to support MAST entrepreneurship. 

\subsection{Societal interventions}
The industry must support colleges in designing curricula, projects, and hands-on experiences for MAST education. Industry should provide financial and personnel support to the colleges in their area to become MAST-enabled colleges. This is in the interest of the industry as well because they get to hire some of these entrepreneurs.  

Working professionals will be required to help implement Faculty Development Programs (FDP) and provide mentorship to the students as they learn and practice to be entrepreneurs. They will become part of the college's advisory committee and help guide the curriculum development and delivering courses. 

We also need to encourage local businesses and individuals to buy local and support these local MAST enterprises and remind them of existing policies and initiatives like "Vocal for Local" \cite{PMIndia2023}. 

\section{Building a supportive ecosystem for MAST enterprises}
\label{sec:mast-enterprises-ecosystem}
While training engineering students to be MAST entrepreneurs is important, building an ecosystem where MAST enterprises can survive and thrive is probably more important. 

When a MAST enterprise is launched, the co-founders need support for all enterprise setup, operations, and growth aspects. 
\begin{enumerate}
    \item \textbf{Social security needs} - To mitigate the risk of failure, the government can offer schemes for life and health insurance, income guarantees, and supporting resources to reduce risk and perception considerably. 
    \item \textbf{Business stability and growth needs} - The government can make it easy for them to run their business by providing a simplified process for launching. Access to inexpensive credits shifts workers from informal microentrepreneurship into formal employment \cite{dehejia2022financial}, which benefits the economy by providing employees to larger firms and moving the microentrepreneurs up the value chain. 
    \item \textbf{Resource needs} - The colleges can easily provide inexpensive/free resources for the MAST enterprises by making it easy for engineering students to do internships and apprenticeships at these enterprises. Interns offer expected and unexpected value to the micro-entrepreneurs \cite{kapasi2023university}, and this can create a positive cycle where interns can learn the skills to be a micro-entrepreneur or how to work in a micro-enterprise. Government can also use a policy like "Internship in Top Companies" \cite{budget2024} to support MAST enterprises.  
    \item \textbf{Mentorship needs} - Government and other ecosystem players can help provide mentorship and career guidance to the MAST entrepreneurs to make key decisions; for example, when to grow, when to exit and join a company, and when to merge with another MAST enterprise. 
\end{enumerate}

\section{Challenges, Risks and Mitigations}
\label{sec:challenges-risks-mitigations}
The proposal above may seem audacious and impractical or too complex to implement. While it is a complex solution that will take a long time, we believe it is achievable if we can get support from various stakeholders. In this section, we identify key challenges to the idea or its implementation and try to address them. These require further debate and discussions, and one of the goals of this position paper is to trigger them. 

\textbf{Graduates who are not good enough to get jobs today are not good enough to be entrepreneurs, so this proposal can't work. }

As argued in previous sections, there aren't enough jobs in the market. Therefore, many capable but jobless graduates will benefit from this proposal. In addition, we believe that the disproportionately high focus on programming ability during the selection process filters out many capable graduates who possess strong skills suitable for entrepreneurship, and such a proposal can help them be MAST entrepreneurs. As long as we can align the students with the skills they can excel in and help them learn entrepreneurial (and other) skills through hands-on experiences, this proposal will be able to make a significant impact on a large number of students. 

\textbf{Many students lack the language or cultural skills, which makes them unfit for jobs. How will they succeed in entrepreneurship?}
Language or culture is important in these jobs because they serve global customers. By being hyperlocal, MAST enterprises do not have these needs - local language and cultural awareness are perfectly fine and, in fact, desirable. Hence, a deficiency in jobs can be a merit in MAST.   

\textbf{The jobs being done by MAST enterprises are jobs current IT companies do, so if MAST is successful, IT jobs will reduce further.}
MAST enterprises will operate in a hyperlocal environment where there are no software, technology, or AI solutions available because it is not practical for regular companies to offer solutions to them. In addition, many of these are expected to be in Tier 2/3 cities with little technology penetration. Hence, we believe these are net new jobs and contributions to the GDP. The training of MAST entrepreneurs ensures they learn the skills to thrive in such an environment. 

\textbf{Entrepreneurship requires risk, capital, and patience; how will so many people handle this, especially given the fact that many of them come from middle-class families and need to earn quickly? }

Typical entrepreneurship is Model A - a high-risk, high-reward, global-play model. The MAST entrepreneurship model, on the other hand, is a low-risk, low-reward, local-play model, what we have called Model B above. Our proposal also includes interventions to address risk and capital requirements through policies like income guarantees, insurance against business failures, and funding through skilling and entrepreneurship programs that already operate. The goal is to make MAST entrepreneurship as easy as a job.

\section{Conclusion and the way forward}
\label{sec:conclusion-futurework}
While the supply of engineering graduates continues to stay very high (900K per year), the jobs for engineering graduates continue to stay much smaller (less than 300K per year even with a very liberal estimate). The headwinds of productivity improvements and AI as an enabler of significant automation mean that such a significant supply-demand gap will likely stay for a long time. 

To mitigate the widespread unemployment of engineers, this paper proposed that we train our engineering students to be micro-entrepreneurs who focus on AI, software, and technology (MAST) skills and deliver product-driven services for hyperlocal problems. These MAST entrepreneurs will run micro-enterprises. We recommend several curricular, policy, and societal interventions to ensure we produce MAST entrepreneurs at scale who can run MAST enterprises after graduation. This can address the huge gap between supply and demand. 

Three key stakeholders need to come together to enable this: a) Colleges must design and deliver an ecosystem where micro-entrepreneurship can be taught, learned, and practiced, b) Policymakers must implement policies that support these entrepreneurs right out of the college in setting up, operating, and growing their micro-enterprises, and c) The society (including industry and professionals) must be willing and able to support this ecosystem through mentorship, coaching and economic activities that encourage local businesses.  

This is an ambitious plan, and it will take time to operationalize. However, we believe that the enormity of the problem requires such an ambitious solution, and academia, society, and policymakers must collaborate and implement this solution. 

We hope this paper will be a trigger for conversations that can lead to collaborative efforts to define, refine, and implement various components of the solution proposed and initiate the journey toward solving this complex social problem at India scale.

\begin{credits}
\subsubsection{\ackname} This paper has benefitted from conversations with many IT industry specialists in India.

\subsubsection{\discintname}
The authors have no competing interests to declare relevant to this article's content. 
\end{credits}
%
%
%
\bibliographystyle{splncs04}
\bibliography{adhoc}

\begin{thebibliography}{10}
\providecommand{\url}[1]{\texttt{#1}}
\providecommand{\urlprefix}{URL }
\providecommand{\doi}[1]{https://doi.org/#1}

\bibitem{aicte2024}
AICTE: All india council for technical education (2024), \url{https://www.aicte-india.org/}

\bibitem{audretsch2012entrepreneurship}
Audretsch, D.: Entrepreneurship research. Management decision  \textbf{50}(5),  755--764 (2012)

\bibitem{da2015entrepreneurship}
Da~Silva, G.B., Costa, H.G., De~Barros, M.D.: Entrepreneurship in engineering education: A literature review. International Journal of Engineering Education  \textbf{31}(6),  1701--1710 (2015)

\bibitem{dehejia2022financial}
Dehejia, R., Gupta, N.: Financial development and micro-entrepreneurship. Journal of Financial and Quantitative Analysis  \textbf{57}(5),  1834--1861 (2022)

\bibitem{IndiaToday2020}
Deka, K.: The coaching game (August 2020), \url{https://tinyurl.com/3spmrf7a}, accessed: 2024-07-30

\bibitem{ericsson2015acquisition}
Ericsson, K.A.: Acquisition and maintenance of medical expertise: a perspective from the expert-performance approach with deliberate practice. Academic Medicine  \textbf{90}(11),  1471--1486 (2015)

\bibitem{forrester1997industrial}
Forrester, J.W.: Industrial dynamics. Journal of the Operational Research Society  \textbf{48}(10),  1037--1041 (1997)

\bibitem{hussain2014entrepreneurship}
Hussain, M.D., Bhuiyan, A.B., Bakar, R.: Entrepreneurship development and poverty alleviation: an empirical review. Journal of Asian scientific research  \textbf{4}(10), ~558 (2014)

\bibitem{kapasi2023university}
Kapasi, I.: How university internships benefit microbusiness owners: Beyond anticipated value. Industry and Higher Education  \textbf{37}(1),  80--93 (2023)

\bibitem{katzenbach2015wisdom}
Katzenbach, J.R., Smith, D.K.: The wisdom of teams: Creating the high-performance organization. Harvard Business Review Press (2015)

\bibitem{marchand2014student}
Marchand, J., Hermens, A.: Student entrepreneurship: a research agenda. In: Australian and New Zealand Academy of Management Conference (2014)

\bibitem{mgimsme2024}
{McKinsey Global Institute}: A microscope on small businesses (May 2024), \url{https://mck.co/3WmubDR}

\bibitem{graduateskillindex2023}
{Mercer | Mettl}: India’s graduate skill index 2023 (2023), \url{https://pages.mettl.com/indias-graduate-skill-index}

\bibitem{budget2024}
{Minister of Finance}: Budget 2024-2025 speech (July 2024), \url{https://bit.ly/3A6kLVj}

\bibitem{economicsurvey2024}
{Ministry of Finance}: Economic survey 2024-2025 (July 2024), \url{https://bit.ly/3WhRLkZ}

\bibitem{aishe2022}
MoE: All india survey on higher education 2021-22 (2022), \url{https://bit.ly/3Lhfrkz}

\bibitem{indiasoftwareproductecosystem2022}
NASSCOM: India’s software product ecosystem - accelerating growth (March 2022), \url{https://bit.ly/4bGwQxL}

\bibitem{indiantechecosystem2024}
NASSCOM: Indian tech ecosystem - highlights in fy2024 (2024), \url{https://bit.ly/4cWh7vv}

\bibitem{NIRF2023}
NIRF: India rankings 2023: Engineering (July 2023), \url{https://www.nirfindia.org/Rankings/2023/EngineeringRanking.html}

\bibitem{pandit2018examining}
Pandit, D., Joshi, M.P., Tiwari, S.R.: Examining entrepreneurial intention in higher education: An exploratory study of college students in india. The Journal of Entrepreneurship  \textbf{27}(1),  25--46 (2018)

\bibitem{PMIndia2023}
{Prime Minister of India}: Pm urges citizens to be vocal for local (November 2023), \url{https://tinyurl.com/2u4j8hx5}, accessed: 2024-07-30

\bibitem{raju1999teaching}
Raju, P., Sankar, C.S.: Teaching real-world issues through case studies. Journal of Engineering Education  \textbf{88}(4),  501--508 (1999)

\bibitem{venkateswarlu2017establishing}
Venkateswarlu, P.: Establishing a ‘centre for engineering experimentation and design simulation’: a step towards restructuring engineering education in india. European Journal of Engineering Education  \textbf{42}(4),  349--367 (2017)

\end{thebibliography}
\end{document}